\documentclass[%
reprint,11pt,a4paper
nofootinbib,
nobibnotes,
 amsmath,amssymb,
 aps,
pra,
floatfix,
titlepage,
nofootinbib
]{revtex4-1}
\usepackage{soul}
\usepackage{mathtools}
\usepackage{graphicx}
\usepackage{dcolumn}
\usepackage{bm}

\usepackage{nth}
\usepackage{relsize}
\usepackage{lmodern}
\usepackage{slantsc}
\usepackage{enumitem}
\graphicspath{{images/}}

\usepackage[font=footnotesize]{caption}
\begin{document}


\title{Layer Communities in Multiplex Networks}


\author{Ta-Chu Kao}
\affiliation{Department of Physics, University of Oxford, Oxford, United Kingdom}

\author{Mason A. Porter}
\affiliation{Department of Mathematics, University of California Los Angeles, Los Angeles, USA}
\affiliation{Mathematical Institute, University of Oxford, Oxford, United Kingdom}
\affiliation{CABDyN Complexity Centre, University of Oxford, Oxford, United Kingdom}

\begin{abstract}

Multiplex networks are a type of multilayer network in which entities are connected to each other via multiple types of connections. We propose a method, based on computing pairwise similarities between layers and then doing community detection, for grouping structurally similar layers in multiplex networks. We illustrate our approach using both synthetic and empirical networks, and we are able to find meaningful groups of layers in both cases. For example, we find that airlines that are based in similar geographic locations tend to be grouped together in an airline multiplex network and that related research areas in physics tend to be grouped together in an multiplex collaboration network.

\end{abstract}


\maketitle
\pagenumbering{arabic}


\section{Introduction} \label{one}

A network is a widely-used representation to describe the connectivity of a complex system. In a network, entities (represented by nodes) are adjacent to each other via edges \cite{newman2010}. The best-studied type of network is a graph, but recently multilayer networks have been used to encode increasingly complicated structures --- such as multiplex networks, interconnected networks, and time-dependent networks \cite{kivela2014multilayer,boccaletti2014structure} --- in a network. In a multilayer network, each entity is represented by a ``physical node'', and the manifestation of a given node in a specific layer (i.e., a node-layer) is a ``state node''.

A \emph{multiplex network} is a special kind of multilayer network in which physical nodes can be adjacent to each other through different types of intralayer edges and a given entity on a layer can be adjacent to itself on another layer through an interlayer edge. It thereby represents networks with multiple types of relations. The study of multilayer networks is perhaps the most active area of network science, and multiplex networks in particular have been used in the study of many biological, social, and technological systems---including cellular interactions \cite{bennett2015detection}, contagions \cite{granell2013dynamical}, social relationships \cite{magnani2013combinatorial}, scientific collaborations \cite{iacovacci2016extracting}, and flight connections \cite{cardillo2013emergence}. 

In many empirical multiplex networks, there are many intralayer edges that occur between the same pairs of entities in multiple layers \cite{de2015structural,szell2010multirelational}, leading to considerable edge overlap. When a lot of edges overlap in a pair of layers, it is likely that those two layers possess many similar structures in their connectivity patterns \cite{de2015structural,bianconi2013statistical,menichetti2014correlations,battiston2014structural,nicosia2015measuring}, and such similarities may be useful for characterizing similarities among multiple types of connections. For example, in a multiplex communication network (e.g., text messages, phone calls, and e-mails), in which the physical nodes represent people and the layers represent different communication media, two people who communicate in one layer may also be likely to communicate in other layers, yielding edge  overlaps \cite{bianconi2013statistical}.

In a prominent study of multiplex networks, Szell et al. examined six types of interactions---friendship, communication, trade, enmity, aggression, and punishment---between 300,000 players in a massively multiplayer online role-playing game (MMORPG) called Pardus \cite{szell2010multirelational}. They found significant edge overlaps among positive interactions (communication, friendship, and trade) and significant edge overlaps among negative interactions (enmity, punishment, and aggression). This is sensible, as players who communicate with each other are likely to be friends, and players who attack each other are likely to be enemies. In other words, positive interactions are likely to possess edge overlaps with each other, and the same is true for negative interactions. Understandably, Szell et al. also found few edge overlaps between positive interactions and negative interactions, illustrating that different types of interactions can sometimes fall into natural groups according to their structural similarities. That is, in Pardus, the six interactions can be divided into a group of positive interactions (friendship, communication, and trade) and a group of negative interactions (enmity, aggression, and punishment). 

The tendency for edge overlaps to occur in a heterogeneous manner that depends on relationship type motivates us to introduce the concept of \emph{layer communities}, a group of structurally similar layers that are structurally dissimilar to other layers. A layer community is a type of mesoscale structure that can occur in a multilayer network, such as a multiplex (i.e., multirelational) network. Studying mesoscale structures in networks can be very insightful, and many different types of such structures have been examined. The best-studied type of mesoscale structure is community structure \cite{porter2009communities,fortunato2010community,santo2016}, and other well-known types of mesoscale structure are core--periphery structure \cite{csermely} and roles and positions \cite{rossi2015}. In contrast to standard community structure, we wish to cluster layers rather than nodes, and most mesoscale structures that have been examined are concerned with clustering nodes. For example, a prototypical community (which we will call a ``node community'') consists of a set of densely connected nodes with sparse connections to other sets of nodes \cite{porter2009communities,fortunato2010community,santo2016}. Therefore, edge densities within node communities tend to be high, and edge densities between node communities tend to be low. One can also cluster edges to study ``edge communities'' \cite{ahn2010}, and in the present paper we cluster layers to study ``layer communities''.

Past studies of layer similarities in multilayer networks have focused primarily on node-characteristic similarities, such as interlayer degree correlations and node-community similarities \cite{battiston2014structural,nicosia2015measuring,min2014network,iacovacci2016extracting}. 
These ideas have yielded insights into phenomena such as the presence and consequences (e.g., on percolation and spreading processes) of nontrivial multiplex correlations in networks \cite{dedom2016,lee2015towards}. For example, Iacovacci et al. used a node-characteristic similarity measure to find layer communities (though without explicitly developing the notion of ``layer communities'' or proposing such terminology) in a collaboration network of publications in physics journals and a multiplex social network in the Department of Computer Science at Aarhus University \cite{magnani2013combinatorial,iacovacci2015mesoscopic,iacovacci2016extracting}.
Reference \citet{de2016spectral} defined a quantum-entropy similarity measure by calculating Jensen--Shannon (JS) divergence between two layers and used their measure to cluster layers in a human microbiome multiplex network. 

To study layer communities, we define a novel measure of interlayer structural similarity measure using calculations of edge overlaps. That is, rather than measuring similarity in node characteristics or similarity in quantum entropy as in previous work, we directly measure similarity in connection patterns. Importantly, our goal is to examine layer similarity rather than layer redundancy, which can be used for aggregating layers in multilayer networks to reduce system size \cite{taylor2016,taylor2016detectability,de2015structural,chen2016multilayer}. We seek to develop a method that can meaningfully classify different types of connections in multilayer networks using measures of layer similarities. Such classification has the potential to help infer commonalities between different types of connections in large networks (e.g., common purpose, physical mechanisms, and constraints), and we successfully demonstrate the utility of our approach using three multiplex networks constructed from empirical data.

The rest of our paper is structured as follows. In Section \ref{two}, we propose a new interlayer similarity measure, called \emph{connection similarity}, which is based on pairwise similarity in connection patterns. We then use this measure to cluster layers in synthetic multiplex networks in Section \ref{3a} and in three empirical multiplex networks in Section \ref{3b}. We conclude in Section \ref{four}.


\section{Connection Similarity}\label{two}

\begin{figure}[h!]
\includegraphics[width=\columnwidth]{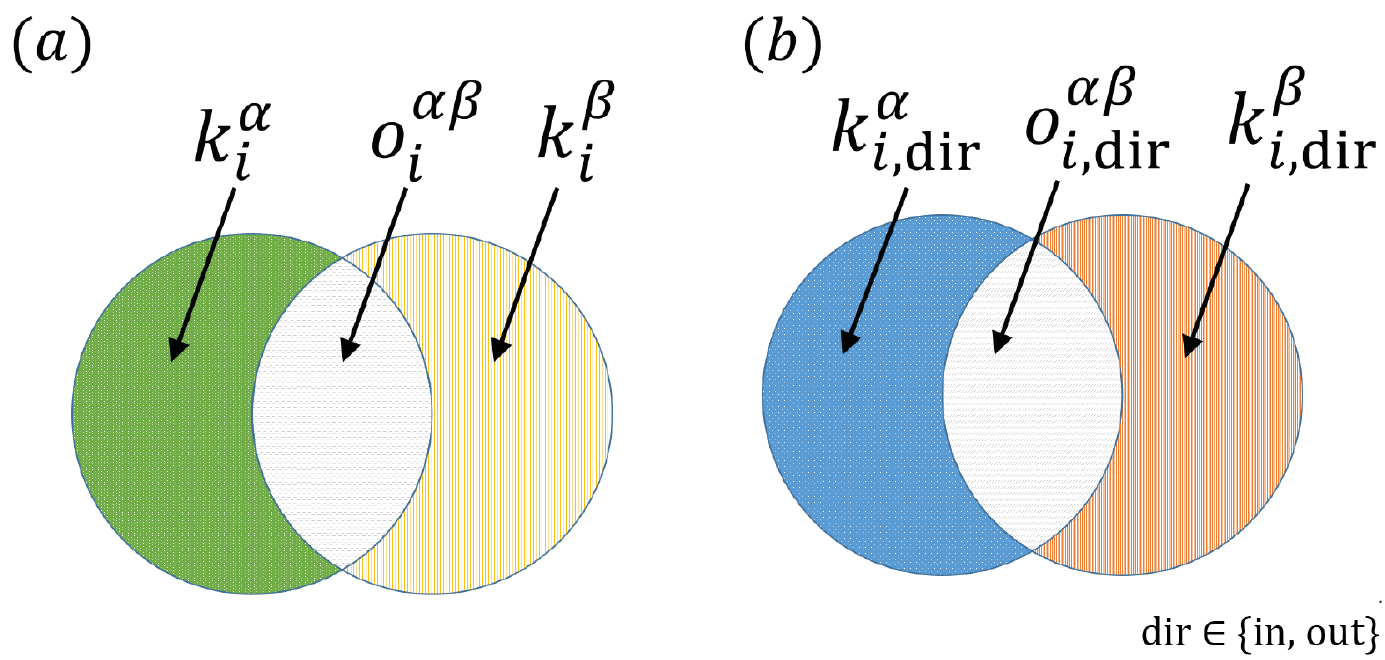}
	\caption{Schematic illustrations of (a) undirected local similarity $\phi^{\alpha\beta}_{i}$ and (b) directed local similarity $\phi^{\alpha\beta}_{i,\text{dir}}$.
	}
	\label{fig1}
\end{figure}

Consider a multiplex network that has $M$ layers and $N$ nodes in each layer, where we assume for simplicity that every node exists on every layer and that there are no interlayer edges (so that we are studying edge-colored multigraphs). Following convention \cite{kivela2014multilayer,de2013mathematical}, we use the Roman alphabet to label nodes and the Greek alphabet to label layers. A multiplex network $\mathcal{G}=\{\mathcal{G}^{1}, \dots ,\mathcal{G}^{\alpha}, \dots ,\mathcal{G}^{M}\}$ without interlayer edges is a set of $M$ monolayer networks, where $\mathcal{G}^{\alpha}$ denotes the monolayer network on layer $\alpha$, which we represent as an $N \times N$ weighted adjacency matrix $W^{\alpha}$. An element $w^{\alpha}_{ij}$ of $W^{\alpha}$ represents the weight of an intralayer edge from node $i$ to node $j$ on layer $\alpha$, where $i,j \in \{1, \dots ,N\}$ and $\alpha \in \{1, \dots ,M\}$.  

A simple way to quantify interlayer similarity is to count the number of edge overlaps between two layers (ignoring the weights of the edges) \cite{bianconi2013statistical,menichetti2014correlations,battiston2014structural,nicosia2015measuring,de2015structural}. 
There is an overlapping edge between nodes $i$ and $j$ in layers $\alpha$ and $\beta$ if and only if there is an edge between nodes $i$ and $j$ in both $\alpha$ and $\beta$ (i.e., $\theta(w^{\alpha}_{ij})=1$ and $\theta(w^{\beta}_{ij})=1$), where $\theta(x) = 1$ if $x > 0$ and $\theta(x) = 0$ otherwise.

We consider the \emph{local overlap} \cite{cellai2013percolation,cellai2016message} 
\begin{equation*}
	o^{\alpha\beta}_{i}  = \sum\limits_{j} \theta(w^{\alpha}_{ij})\theta(w^{\beta}_{ij}) \,,
\end{equation*}	
which counts the number of overlapping edges that are incident to node $i$ in both layer $\alpha$ and layer $\beta$. In an undirected multiplex network, the local overlap $o^{\alpha\beta}_{i}$ quantifies the similarity between the connection patterns of node $i$ in layer $\alpha$ and node $i$ in layer $\beta$.

The local overlap $o^{\alpha\beta}_{i}$ does not account for the intralayer degrees of node $i$ in layers $\alpha$ and $\beta$, even though degree contributes to the total number of overlapping edges. To take degree into account in an undirected multiplex network, we define \emph{local similarity} 
\begin{equation}
	\phi^{\alpha\beta}_{i} = \frac{o_{i}^{\alpha\beta}}{k_{i}^{\alpha}+k_{i}^{\beta} - o_{i}^{\alpha\beta}} \in [0,1]\,, 
\end{equation}	
where $k^{\alpha}_{i} = \sum_{j} \theta(w^{\alpha}_{ij})$ is the degree of node $i$ in layer $\alpha$. Local similarity $\phi^{\alpha\beta}_{i}$ calculates the number of overlapping edges that are incident to node $i$ in layers $\alpha$ and $\beta$ as a proportion of the number of unique edges that are incident to node $i$ in the two layers (i.e., $k_{i}^{\alpha}+k_{i}^{\beta} - o_{i}^{\alpha\beta}$) [see Fig.~\ref{fig1}(a)]. The local similarity $\phi^{\alpha\beta}_{i} = 1$ if and only if all of the edges that are incident to node $i$ in layers $\alpha$ and $\beta$ overlap, and $\phi^{\alpha\beta} = 0$ if and only if none of the edges that are incident to node $i$ in layers $\alpha$ and $\beta$ overlap. 

We then define \emph{connection similarity}
\begin{equation} \label{eq1}
	\phi^{\alpha\beta} = \frac{1}{N} \sum\limits_{i} \phi^{\alpha\beta}_{i} \in [0,1] 
\end{equation} 
to calculate the mean local similarity between layers $\alpha$ and $\beta$ and thereby quantify the similarity between the connection patterns in the two layers.

Thus far, we have considered connection similarity in an undirected multiplex network, but it is straightforward to generalize this notion to directed multiplex networks. First, we distinguish between the number of overlapping edges that are connected \emph{to} node $i$ in layers $\alpha$ and $\beta$ [specifically, we calculate $o^{\alpha\beta}_{i,\text{in}} = \sum\limits_{j}\theta(w^{\alpha}_{ij})\theta(w^{\beta}_{ij})$] and the number of overlapping edges that are connected \emph{from} node $i$ in layers $\alpha$ and $\beta$ [specifically, we calculate $o^{\alpha\beta}_{i,\text{out}} = \sum\limits_{l}\theta(w^{\alpha}_{li})\theta(w^{\beta}_{li})]$. We also need to distinguish between in-degree $k^{\alpha}_{i,\text{in}} = \sum_{j} \theta(w^{\alpha}_{ji})$ and out-degree $k^{\alpha}_{i,\text{out}} = \sum_{l} \theta(w^{\alpha}_{il})$. 

We define connection similarity in a directed multiplex network as
\begin{equation} \label{eq2}
	 \phi^{\alpha\beta} = \frac{1}{2N} \sum_{i} (\phi^{\alpha\beta}_{i,\text{in}} + \phi^{\alpha\beta}_{i,\text{out}}) \,,
\end{equation} 
where 
\begin{equation}
	\phi^{\alpha\beta}_{i,\text{dir}} = \frac{o_{i,\text{dir}}^{\alpha\beta}}{k_{i,\text{dir}}^{\alpha}+k_{i,\text{dir}}^{\beta} - o_{i,\text{dir}}^{\alpha\beta}}
\end{equation}	
for $\text{dir} \in \{\text{in},\text{out}\}$ [see Fig.~\ref{fig1}(b)]. Equations \eqref{eq1} and \eqref{eq2} are equivalent in an undirected multiplex network, because $w^{\alpha}_{ij} = w^{\alpha}_{ji}$ in that case.


V\"{o}r\"{o}s et al. recently defined a layer similarity measure similar to connection similarity \cite{voros2017cluster}. Their similarity measure is
\begin{equation}
	J^{\alpha\beta} = \frac{n^{\alpha\beta}_{11}}{n^{\alpha\beta}_{11} + n^{\alpha\beta}_{10} + n^{\alpha\beta}}\,,
\end{equation} %
where 
\begin{widetext}
\begin{equation}
	n^{\alpha\beta}_{mn} = \sum_{i<j} I[\theta(w^{\alpha}_{ij}) = m, \theta(w^{\alpha}_{ij}) = n] \,, \qquad m,n \in \{0,1\}\,, \quad \alpha,\beta \in \{1 \dots M \}\,,
\end{equation}
\end{widetext}
where $I[A]$ is the indicator function of the set $A$. Rewriting their similarity measure using our notation yields
\begin{equation}
	J^{\alpha\beta} = \frac{O^{\alpha\beta}}{m^{\alpha} + m^{\beta} - O^{\alpha\beta}}\,,
\end{equation} %
where $O^{\alpha\beta} = \sum_{i<j} \theta(w^{\alpha}_{ij})\theta(w^{\beta}_{ij})$ is the global edge overlap and $m^{\alpha}$ is the total number of edges on layer $\alpha$. The quantity $J^{\alpha\beta}$ is a Jaccard similarity between layers $\alpha$ and $\beta$.

In comparison to Jaccard similarity, connection similarity puts more emphasis on local overlap than global overlap. In \cite{voros2017cluster}, V\"{o}r\"{o}s et al. used their similarity measure to cluster layers in a high-school social network and thereby reduce system size. In Section \ref{3e}, we compare the layer communities that we find using connection similarity with their Jaccard similarity measure. 


\section{Detection of Layer Communities} \label{three}

To find layer communities in a multiplex network $\mathcal{G}$, we create a monolayer network $G_L$ with adjacency matrix ${\bf A}$ in which the nodes are the layers in $\mathcal{G}$ and the edge weights are the interlayer similarities between the layers in $\mathcal{G}$. One can then detect node communities in $G_L$ using any of the myriad available methods \cite{santo2016}. In this paper, we use the Louvain method \cite{blondel2008fast} and InfoMap \cite{rosvall2008maps,MapCode} on $G$ to find layer communities of a multiplex network $\mathcal{G}$. We examine both synthetic networks and empirical networks.
 
Iacovacci et al. also constructed a similarity network from a multiplex network to cluster layers \cite{iacovacci2015mesoscopic,iacovacci2016extracting}, but they used an interlayer node-similarity measure rather than connection similarities. To define a measure of layer similarity, they used the idea of a network ensemble \cite{bianconi2007entropy,bianconi2009entropy}. A network ensemble (i.e., a probability distribution on networks) is a set of possible networks that satisfy structural constraints, such as certain node properties (e.g., the degree or community assignment in a specified layer) and the probability of drawing each network from the collection. Let $q^{\alpha}_{i} \in \{ 1, \dots, Q^{\alpha}\}$ (where $Q^{\alpha}$ denotes the maximum value of the property) denote some property of node $i$ in layer $\alpha$. Given some property $q^{\alpha}_{i}$, Iacovacci et al. defined the class $c^{\alpha}_{i} = f(k^{\alpha}_{i},q^{\alpha}_{i}) \in \{1, \dots ,C^{\alpha} \}$ (where $C^{\alpha}$ denotes the total number of classes in layer $\alpha$) of a node for some function $f$. They then defined the entropy of layer $\alpha$ with respect to node property $q^{\alpha}_{i}$ as
\begin{equation}
	\Sigma_{k^{\alpha}_{\bullet},q^{\alpha}_{\bullet}} = \log \Bigg [\prod\limits_{c < c'}{n^{\alpha}_{c}n^{\alpha}_{c'} \choose e^{\alpha}_{c,c'}}\prod\limits_{c} {n^{\alpha}_{c}(n^{\alpha}_{c'}-1)/2 \choose e^{\alpha}_{c,c}} \Bigg ] \,,
\end{equation}
where $e^{\alpha}_{c,c'}$ is the number of edges between nodes in class $c$ and nodes in class $c'$. The entropy $ \Sigma_{k^{\alpha}_{\bullet},q^{\alpha}_{\bullet}}$ measures the amount of information in layer $\alpha$ with respect to property $q_{i}^{\alpha}$.  They then calculated a z-score 
\begin{equation}
	\Theta_{k^{\alpha}_{\bullet},q^{\beta}_{\bullet}} = \frac{E_{\pi} [\Sigma_{k^{\alpha}_{\bullet},\pi(q^{\beta}_{\bullet})}] - \Sigma_{k^{\alpha}_{\bullet},q^{\alpha}_{\bullet}}}  {\sigma_{\pi} [\Sigma_{k^{\alpha}_{\bullet},\pi(q^{\beta}_{\bullet})}]} 
\end{equation}	
to quantify the amount of information on layer $\alpha$ relative to a uniformly random permutation $\pi$ of node properties $q^{\beta}_{i}$ on layer $\beta$. Here, $E_{\pi} [\Sigma_{k^{\alpha}_{\bullet},q^{\alpha}_{\bullet}}]$ is the expected entropy and $\sigma_{\pi} [\Sigma_{k^{\alpha}_{\bullet},q^{\beta}_{\bullet}}]$ is the standard deviation over the random permutation $\pi$. Finally, Iacovacci et al. defined a symmetric indicator function 
\begin{equation}
	\Theta^{\text{S}}_{\alpha\beta} = \frac{1}{2} \big( \frac{\Theta_{k^{\alpha}_{\bullet},q^{\beta}_{\bullet}}}{\Theta_{k^{\alpha}_{\bullet},q^{\alpha}_{\bullet}}} + \frac{\Theta_{k^{\beta}_{\bullet},q^{\alpha}_{\bullet}}}{\Theta_{k^{\beta}_{\bullet},q^{\beta}_{\bullet}}} \big)
\end{equation}
to quantify the similarity between layers $\alpha$ and $\beta$ with respect to property $q^{\alpha}_{i}$. In this article, we refer to the indicator function $\Theta^{\text{S}}_{\alpha\beta}$ as the \emph{mesoscopic similarity} between layers $\alpha$ and $\beta$. 

The crucial difference between our approach and that of Iacovacci et al. \cite{iacovacci2015mesoscopic,iacovacci2016extracting} is that we measure the connection similarity (an edge-centric property) between two layers instead of a similarity in their node properties. We also calculate layer similarity based on a measure of edge overlaps instead of using an explicitly information-theoretic approach. In Section \ref{3d}, we compare the layer communities that we find using our approach and the approach of Iacovacci et al. in a network constructed from empirical data.

Domenico et al. proposed a layer similarity measure that quantifies the Jensen--Shannon (JS) distance between the Von Neumann entropies of two layers \cite{de2015structural}. They defined the Von Neumann entropy of a layer $\alpha$ as
\begin{equation} 
	h(w^{\alpha}_{ij}) = -\text{Tr}(L^{\alpha}\log L^{\alpha}) \,,
\end{equation}	
where 
\begin{equation}
	L^{\alpha}_{ij} = \frac{1}{\sum_{i<j} \theta(w^{\alpha}_{ij})} \; \bigg[\text{diag}(k^{\alpha}_{i}) - \theta(w^{\alpha}_{ij}) \bigg]
\end{equation}	
is an element of $L^{\alpha}$. They defined the JS distance between layers $\alpha$ and $\beta$ as
\begin{widetext}
\begin{equation}
	D_{\text{JS}}^{\alpha\beta} = \sqrt{h\left(\frac{1}{2}\left[w^{\alpha}_{ij} + w^{\beta}_{ij}\right]\right) - \frac{1}{2}\left[h\left(w^{\alpha}_{ij}) + h(w^{\beta}_{ij}\right)\right]} \in [0,1]\,.
\end{equation}
\end{widetext}
Domenico et al. showed that $1 - D_{\text{JS}}^{\alpha\beta}$ can be used to quantify similarity between layers $\alpha$ and $\beta$. They used a quality function based on such a measure to cluster layers and thereby reduce the number of layers in a multilayer network. In our subsequent discussions, we refer to the quantity $1 - D_{\text{JS}}^{\alpha\beta}$ as the \emph{JS similarity} between layers $\alpha$ and $\beta$. Instead of focusing on the difference in information contained in the two layers, our goal is to directly compare the connection patterns of pairs of layers. In Section \ref{3d}, we compare the layer communities that we find using the connection similarity measure and the JS similarity measure. 


\subsection{Layer Communities in Benchmark Networks}\label{3a}

\begin{figure}[h]
\includegraphics[width=\columnwidth]{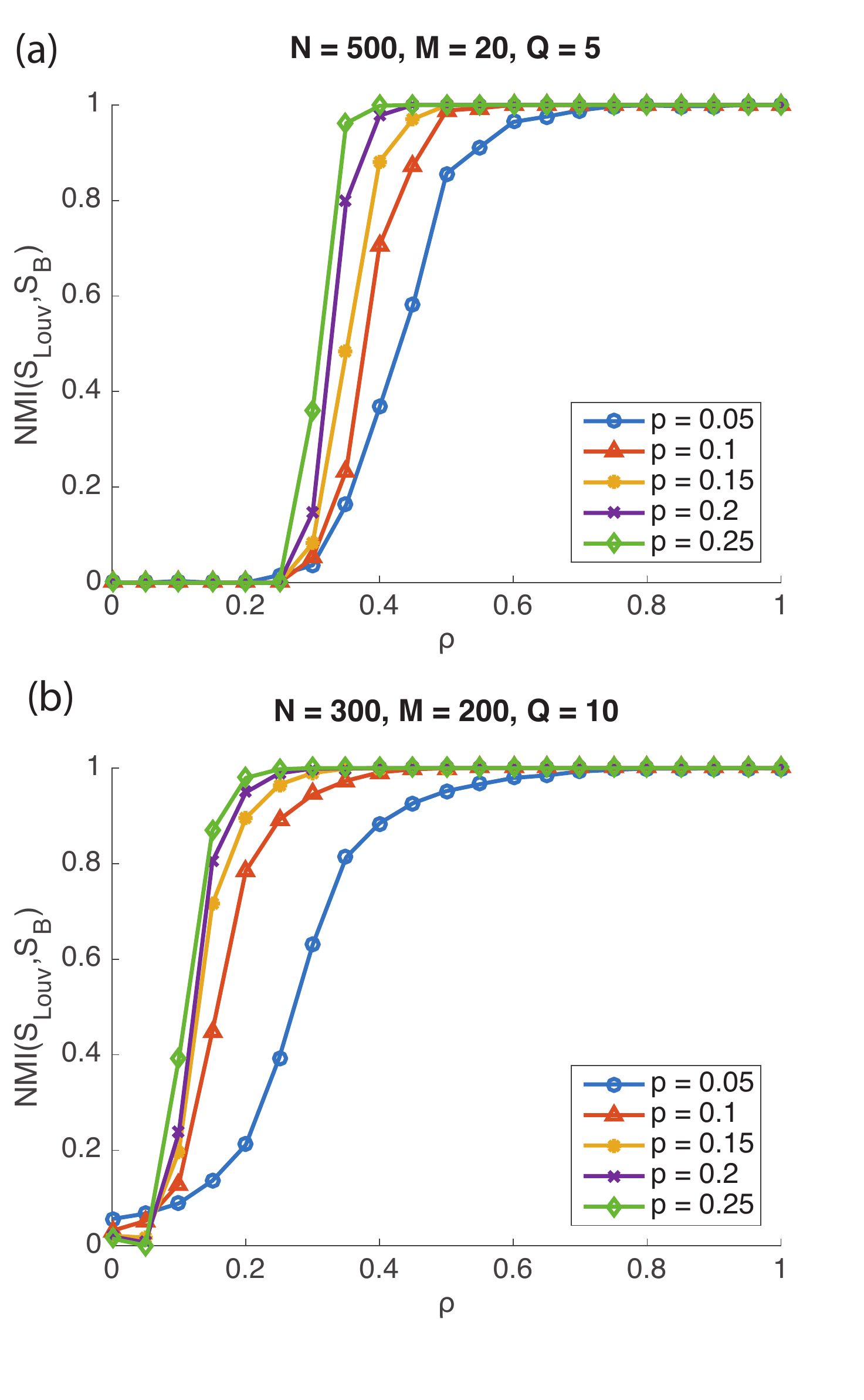}
	\caption{Normalized mutual information (NMI) between the layer communities that we obtain using the Louvain method \cite{blondel2008fast} and planted layer communities for different values of the probability correlation $\rho$ and the mean probability $p$ that two nodes are adjacent. The benchmark multiplex networks in panel (a) have $500$ nodes, $20$ layers, and $5$ layer communities. The benchmark multiplex networks in panel (b) have $300$ nodes, $200$ layers, and $10$ layer communities. Each data point is a mean over $100$ simulations.
	}
	 \label{fig2}
\end{figure}

To test our approach, we construct multiplex benchmark networks with $M$ layers, $N$ nodes in each layer, and $Q$ planted layer communities. The planted layer community assignment is indicated by the vector $\mathbf{S}_{\text{B}}$, where $S^{\alpha}_{\text{B}} \in \{1, \dots ,Q\}$ is the planted layer community of layer $\alpha$. 

To create one of these benchmark networks, we connect nodes $i$ and $j$ on layer $\alpha$ with probability $p^{\alpha}_{ij} \in [0,1]$. In other words, for each $i$ and $j$ (with $i \neq j$), we set $\theta(w^{\alpha}_{ij}) = 1$ with probability $p^{\alpha}_{ij}$ and $\theta(w^{\alpha}_{ij}) = 0$ with probability $1-p^{\alpha}_{ij}$. To introduce interlayer similarity into these benchmarks, we sample $p^{\alpha}_{ij}$ from a multivariate Gaussian copula. The Gaussian copula is a distribution over the cube $[0,2p]^{MN(N-1)/2}$, where $p \in [0,0.5]$. In other words, $p^{\alpha\beta}_{ij}$ is uniformly distributed between $0$ and $2p$, where $p$ is the mean probability that two nodes are adjacent. We construct the copula's correlation matrix so that $p^{\alpha}_{ij}$ and $p^{\beta}_{ij}$ are positively correlated if and only if layers $\alpha$ and $\beta$ are in the same layer community.  Specifically, the correlation between $p^{\alpha}_{ij}$ and $p^{\beta}_{ij}$ is $\rho \in [0,1]$, where $\rho > 0$ if $S^{\alpha}_{\text{B}} = S^{\beta}_{\text{B}}$ and $\rho = 0$ otherwise. We henceforth use the term ``probability correlation'' for $\rho$.

We distinguish our notation for the layer communities that we find using the Louvain method \cite{blondel2008fast} from the layer communities that we find using InfoMap \cite{rosvall2008maps,MapCode} by writing the former as $\mathbf{S}_{\text{Louv}}$ and the latter as $\mathbf{S}_{\text{Info}}$. 

A community assignment is a vector whose components indicate the community of each node. To compare two community assignments $\mathbf{X}$ and $\mathbf{Y}$, we calculate normalized mutual information (NMI) \cite{danon2006effect,strehl2002} between them:
\begin{equation}
	\text{NMI}(\mathbf{X},\mathbf{Y}) = \frac{H(\mathbf{X}) + H(\mathbf{Y}) - H(\mathbf{X},\mathbf{Y})}{H(\mathbf{X})H(\mathbf{Y})}\,, 
\end{equation}	
where $H(\mathbf{X})$ is the Shannon entropy of community assignment $\mathbf{X}$ and $H(\mathbf{X},\mathbf{Y})$ is the joint Shannon entropy of community assignments $\mathbf{X}$ and $\mathbf{Y}$.  

When $\text{NMI}(\mathbf{X},\mathbf{Y}) = 1$, the two layer community assignments $\mathbf{X}$ and $\mathbf{Y}$ are equivalent. That is, $X^{\alpha} = X^{\beta}$ if and only if $Y^{\alpha} = Y^{\beta}$ and $X^{\alpha} \neq X^{\beta}$ if and only if $Y^{\alpha} \neq Y^{\beta}$. When $\text{NMI}(\mathbf{X},\mathbf{Y}) = 0$, the two layer community assignments $\mathbf{X}$ and $\mathbf{Y}$ are independent of each other.

In Fig.~\ref{fig2}, we show that as the correlation $\rho$ increases, there is a sigmoid-like transition in $\text{NMI}(\mathbf{S}_{\text{Louv}},\mathbf{S}_{\text{B}})$ from $0$ to $1$.  This suggests that our method is able to detect the planted layer communities when the correlation $\rho$ is above some threshold. However, we find that InfoMap \cite{rosvall2008maps,MapCode} clusters all of the layers into the same layer community. Hence, $\text{NMI}(\mathbf{S}_{\text{Info}}\mathbf{S}_{\text{B}}) = 0$ for all correlations $\rho$. This suggests that InfoMap \cite{rosvall2008maps,MapCode} is unable to find the correct planted layer communities and our approach gives different results for different node-community detection methods. In contrast, the Louvain method \cite{blondel2008fast} is able to detect the planted layer communities when $\rho$ is above a certain threshold.

We also find (see Fig.~\ref{fig2}) that the sigmoid-like transition becomes delayed and progressively more gradual as the probability $p$ decreases from $0.5$ to $0.1$. This result is reasonable, because the width of the Gaussian copula decreases as $p$ decreases. Thus, for the same amount of correlation, layers in different layer communities are less dissimilar at $p = 0.1$ than they are at $p = 0.2$. 


\subsection{Layer Communities in Empirical Multiplex Networks} \label{3b}

We now demonstrate that our approach is able to detect meaningful layer communities in empirical multiplex networks.


\subsubsection{Sampson Monastery Multiplex Social Network}

\begin{figure*}[ht!]
\includegraphics[width=\textwidth]{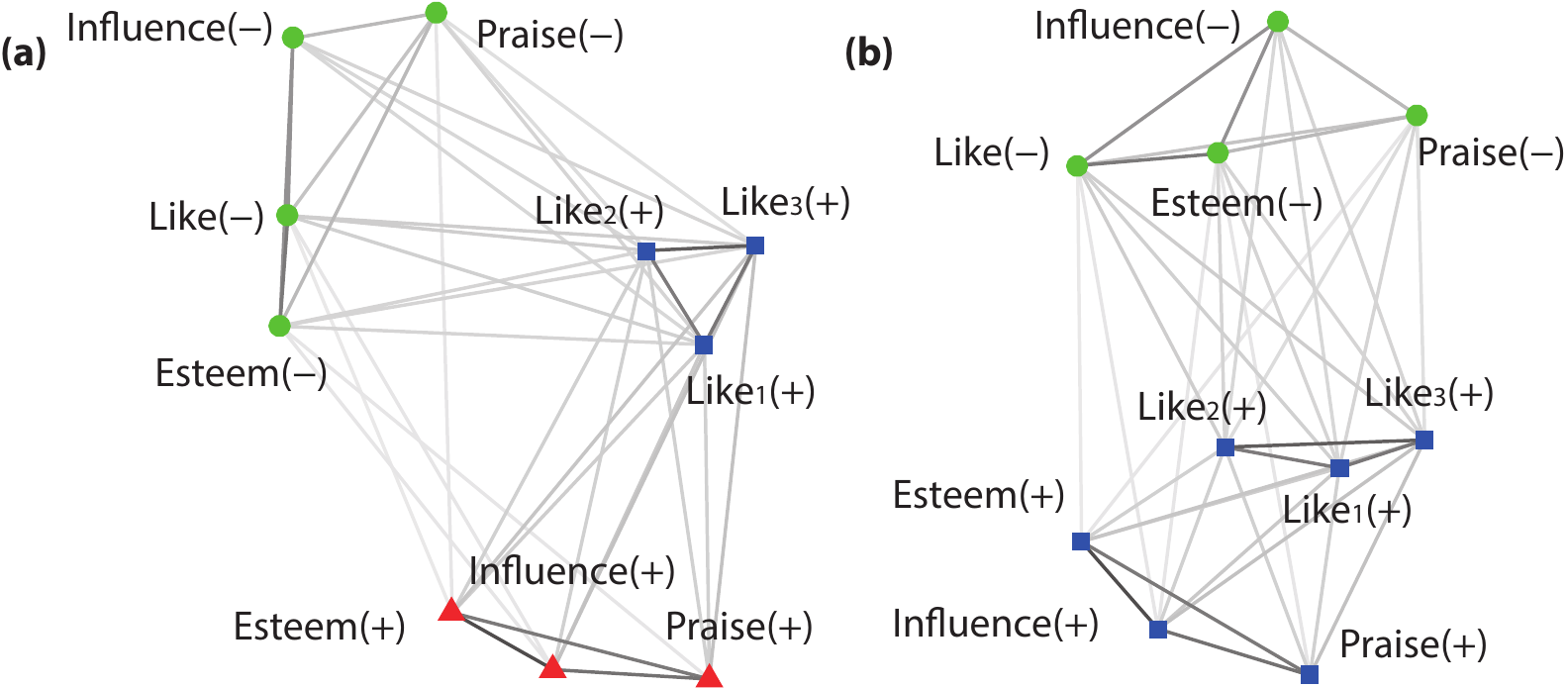}
	\caption{Layer communities in the Sampson monastery data set using (a)  the Louvain methods \cite{blondel2008fast} and (b) InfoMap \cite{rosvall2008maps,MapCode}. We plot the figures the {\sc SpringVisCom} algorithm \cite{JeubSpring,kamada1989algorithm} from \cite{jeub2015think}. We color-code the layer communities and use shapes to represent different layer communities.}
	\label{fig3}
\end{figure*}

In the 1960s, Sampson recorded eight types of relational ties between 18 members in an isolated American monastery for 12 months. The eight relations ties are the following: Like, Esteem, Influence, and Praise; and their negative counterparts\footnote{In his original paper, Sampson used the terms Affect, Esteem, Sanctioning, and Influence (and their counterparts). We use the the label ``Praise($-$)'' and ``Like($+$)" instead of ``Sanctioning'' and ``Affect" in this article.
} \cite{sampson1969crisis,boyd1982social}. Sampson asked each respondent to rank the top-three members for each type of relational tie---e.g., ``List in order those three brothers whom you most esteemed" and ``List in order three brothers whom you esteemed least" \cite{sampson1969crisis,scott2002social,boyd1982social,breiger1975algorithm}. Following \citet{boyd1982social}, we label the eight relational ties as follows: Like($+$), Like($-$), Esteem($+$), Esteem($-$), Influence($+$), Influence($-$), Praise($+$), and Praise($-$). 

Most of the data were collected after several members were expelled from the monastery --- with the exception of Like(+), which was collected in three stages. We use the labels $\text{Like}_{1}\text{($+$)}$, $\text{Like}_{2}\text{($+$)}$, and $\text{Like}_{3}\text{($+$)}$, where $\text{Like}_{1}\text{($+$)}$ and $\text{Like}_{2}\text{($+$)}$ were collected before the expulsion, but $\text{Like}_{3}\text{($+$)}$ was collected after the expulsion. Using data provided by \citet{FreemanData}, we construct a monastery multiplex network with $M=8$ layers and $N=18$ nodes in each layer. Each node represents a member of the monastery, intralayer edges represent relational ties between members, and different layers represent different types of relational ties.

In Fig.~\ref{fig3}(a), we show the layer communities that we obtain using the Louvain method \cite{blondel2008fast}. The negative relational ties are assigned to the same layer community, and except for Like(+), all of the positive relational ties are assigned to the same community. This suggests that the connectivity patterns of $\text{Like}_{1}\text{($+$)}$, $\text{Like}_{2}\text{($+$)}$, and $\text{Like}_{3}\text{($+$)}$ are structurally more similar to each other than they are to those of the other positive relational ties. This is reasonable, because $\text{Like}_{1}\text{($+$)}$, $\text{Like}_{2}\text{($+$)}$, and $\text{Like}_{3}\text{($+$)}$ describe the same relational tie at different times. However, this result appears to differ from a prior observation that $\text{Like}_{1}\text{($+$)}$, $\text{Like}_{2}\text{($+$)}$, and $\text{Like}_{3}\text{($+$)}$ reflect a change in group sentiment over time \cite{FreemanData}.

In Fig.~\ref{fig3}(b), we show the layer communities that we obtain using InfoMap \cite{rosvall2008maps,MapCode}. We find that the negative relational ties are assigned to one layer community and the positive relational ties are assigned to another layer community. Similar to the positive and negative interactions in Pardus (see Section \ref{one}), the negative relational ties are structurally similar, and the positive relational ties --- Like(+), Esteem(+), Praise(+), and Influence(+) --- are structurally similar. This result is consistent with \citet{boyd1982social}'s findings that positive relational ties are highly correlated with each and negative relational ties are highly correlated with each other. To obtain this insight, \citet{boyd1982social} calculated a Pearson correlation between the elements in the weighted adjacency matrices of different layers. (They flattened the two matrices into vectors and then calculated the Pearson correlation between the vectors.)


\subsubsection{Airline Network}

\begin{figure*}[ht!]
\includegraphics[width=\textwidth]{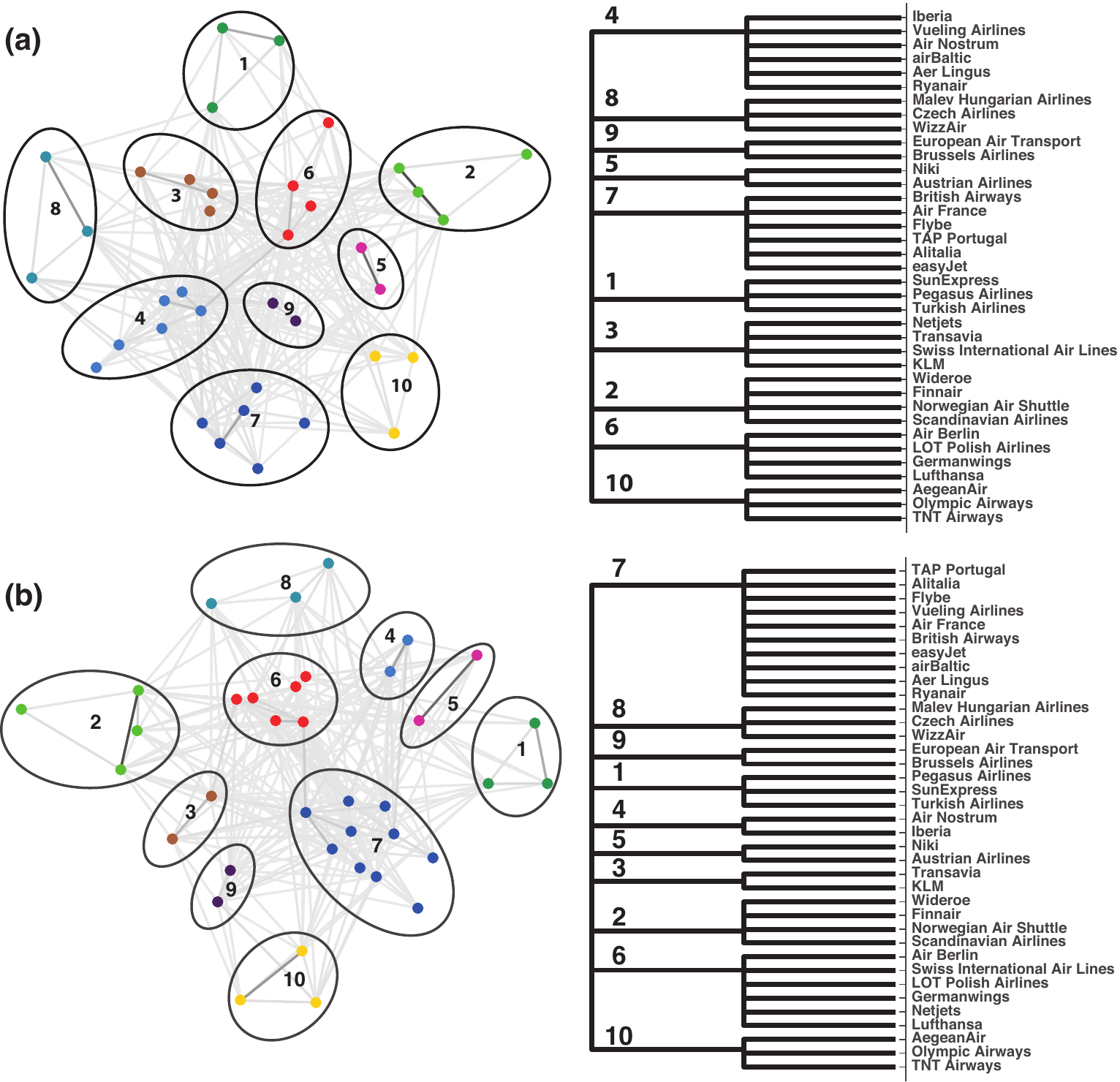}
	\caption{Layer communities in the airline network using (a) the Louvain method \cite{blondel2008fast} and (b) InfoMap. We plot the figures on the left using the {\sc SpringVisComm} algorithm \cite{JeubSpring,kamada1989algorithm}. We color-code and number the layer communities.
	}
	\label{fig4}
\end{figure*}

We construct a multiplex airline network using data compiled by \citet{cardillo2013emergence}. The data set includes the flight connections between 450 airports for 37 different airlines. All of the airports in the data set are located in countries that are part of the European Union (at the time of data collection in 2011). The airline multiplex network has $M=37$ layers and $N=450$ nodes in each layer, although nodes do not have intralayer edges on all layers. The network is undirected and unweighted. Each layer represents a different airline, each node in a layer represents an airport, and each intralayer edge represents an airline-specific flight connection between two airports. 

In Fig.~\ref{fig4}(a), we show the layer communities that we obtain using the Louvain method \cite{blondel2008fast}. The Louvain method \cite{blondel2008fast} partitions the $37$ airlines into $10$ airline communities, and airlines based in the same country or in a similar geographic region tend to be assigned to the same layer community. For example, community 1 consists of all airlines that are based in Turkey, community 7 includes all airlines that are based in Belgium, and community 5 includes all airlines that are based in Scandinavian countries. In Fig.~\ref{fig4}(b), we show the airline communities that we obtain using InfoMap \cite{rosvall2008maps,MapCode}. Airlines that are based in the same country or a similar geographic region again tend to be assigned to the same layer community. 

\begin{figure}[h!]
\includegraphics[width= 0.95 \columnwidth]{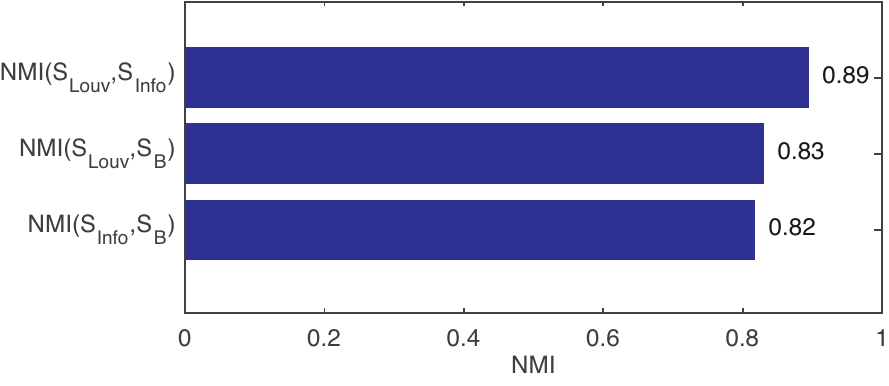}
	\caption{Pairwise NMI between $\mathbf{S}_{\text{Louv}}$, $\mathbf{S}_{\text{Info}}$, and $\mathbf{S}_{\text{B}}$ in the airline network.
	}
	 \label{fig5}
\end{figure}

To build on the above observations, we construct a benchmark community assignment $\mathbf{S}_{\text{B}}$ such that $S^{\alpha}_{\text{B}} = S^{\beta}_{\text{B}}$ if and only if airlines $\alpha$ and $\beta$ are based in the same country or in the same geographic region. More specifically, we assign Wideroe, Finnair, Norwegian Air Shuttle, and Scandinavian Airlines to one layer community because they are all based in Scandinavian countries, and we assign each of the other airlines to a layer community that corresponds to the country in which they are based. We calculate NMI between $\mathbf{S}_{\text{Louv}}$, $\mathbf{S}_{\text{Info}}$, and $\mathbf{S}_{\text{B}}$. In Fig.~\ref{fig5}, we show that $\text{NMI}(\mathbf{S}_{\text{Louv}},\mathbf{S}_{\text{B}})$ and $\text{NMI}(\mathbf{S}_{\text{Info}},\mathbf{S}_{\text{B}})$ are both above $0.8$. The result indicates that the airline communities correspond roughly to the countries or geographic regions in which the airlines are based. We obtain $\text{NMI}(\mathbf{S}_{\text{Louv}},\mathbf{S}_{\text{Info}}) \approx 0.8942$, so the two community assignments are similar. In fact, many airlines are assigned to exactly the same layer community. For example, communities 1, 2 5, 8, 9, and 10 in Fig.~\ref{fig4}(a) are identical to those in Fig.~\ref{fig4}(b). 

Our results are consistent with past research. Reference \citet{cardillo2013emergence} found that major airlines largely follow a hub-and-spoke structure, as there are a few airport hubs in the major cities of a country and many smaller airports scattered around the country that connect to these hubs. This kind of structure allows major airlines---and, in particular, national airlines---to cover an entire country or geographic region \cite{cardillo2013emergence,barthelemy2011spatial,bryan1999hub}. Following a hub-and-spoke structure, airlines that primarily serve the same country or region tend to choose similar large cities in which to set up hubs to connect to remote airports. Consequently, one expects them to have large overlapping connections centered around these common hubs.

\citet{nicosia2015measuring} reported that (due to competition) there is a small overlap in activity pattern of airlines operating in the same region. Moreover, traditional airlines such as Lufthansa tend to have a large overlap in activity pattern with other airlines, whereas low-cost airlines such as easyJet tends to avoid such overlaps. \citet{de2015structural} reported that their algorithm was unable to substantially reduce the number of layers in the airline multiplex network from \cite{cardillo2013emergence} via aggregation of layers that are similar in structure based on JS similarity. \citet{de2015structural} reasoned that airlines tend to minimize edge overlaps to avoid competition. Our results show that airlines operating primarily in the same region tend to have more edge overlaps than airlines that operate primarily in different regions. We are also able to identify airlines that operate in similar regions by grouping them into layer communities.

\subsubsection{American Physical Society (APS) Collaboration Network}

\begin{figure*}[h]
\includegraphics[width=0.9\textwidth]{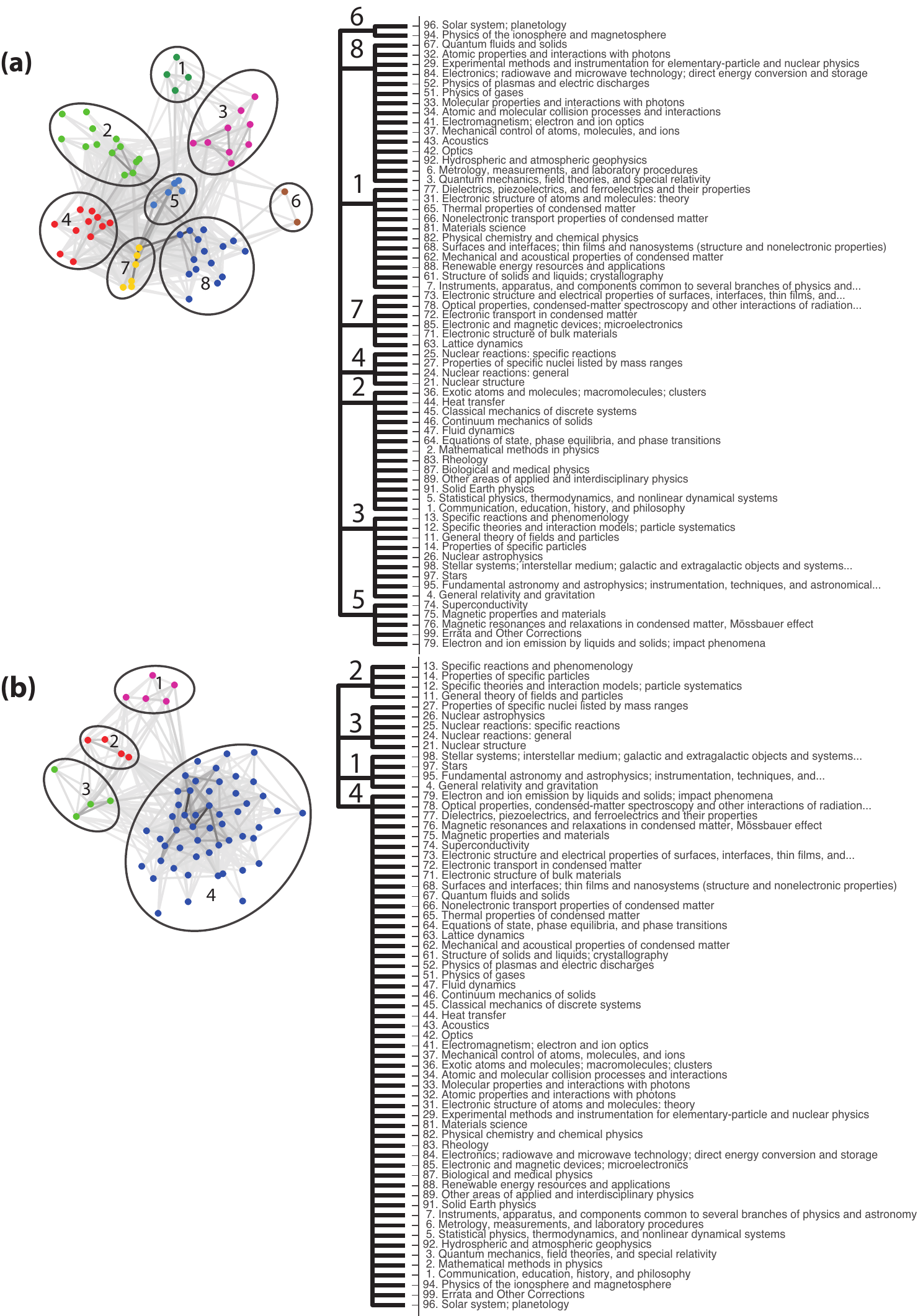}
	\caption{Layer communities in a physics collaboration network. We detect layer communities using (a) the Louvain method \cite{blondel2008fast} and (b) InfoMap \cite{rosvall2008maps,MapCode}. We plot the figures on the left using the {\sc SpringVisCom} algorithm \cite{JeubSpring,kamada1989algorithm}.
	}
	\label{fig6}
\end{figure*}

The American Institute of Physics developed the Physics and Astronomy Classification Scheme (PACS) to identify fields and subfields of physics in journals such as the APS journals. PACS codes are divided into sections (e.g., ``10. The Physics of Elementary Particles and Field") and subsections (e.g., ``11. General Theory of fields and particles" and ``12. Specific theories and interaction models; particle systematics"). 

We construct a multiplex APS Collaboration network ($N = 2598$ nodes and $M = 65$ edges) from an APS journal data set \cite{APSData}. We include papers that are coauthored by 10 or fewer people and are published between 2010 and 2014. Each layer in the multiplex network represents a PACS subsection (e.g., ``21. Nuclear Structure"). A node in a layer represents an author, and there is an edge between two authors in a layer if and only if they have coauthored a paper that is classified in the PACS subsection corresponding to that layer. Each person exists on every layer, but nodes do not possess an intralayer edge on all layers.

In Fig.~\ref{fig6}(a), we show the PACS-subsection layer communities that we obtain using the Louvain method \cite{blondel2008fast}. As expected, these layer communities correspond to related research areas in physics. For example, community 5 includes PACS subsections associated with nuclear physics, and community 7 includes PACS subsections associated with astrophysics. This result is consistent with Iacovacci et al.'s finding that layers related to condensed-matter physics and interdisciplinary physics are assigned to the same layer community using their algorithm \cite{iacovacci2016extracting}.

In Fig.~\ref{fig6}(b), we show the PACS-subsection layer communities that we obtain using InfoMap \cite{rosvall2008maps,MapCode}. Community 4 in Fig.~\ref{fig6}(a) is nearly identical to community 3 in Fig.~\ref{fig6}(b), with the exception that ``26. Nuclear Astrophysics" is in the latter but not in the former. Additionally, community 3 in Fig.~\ref{fig6}(a) is a combination of communities 1 and 2 in Fig.~\ref{fig6}(b). These results suggest that both the Louvain method and InfoMap are able identify structural similarities in these layers. The NMI between these partitions is $\text{NMI}(\mathbf{S}_{\text{Louv}},\mathbf{S}_{\text{Info}}) \approx 0.4968$, suggesting that the layer community assignments are similar (though far from identical).

\begin{figure}[h!]
\includegraphics[width=\columnwidth]{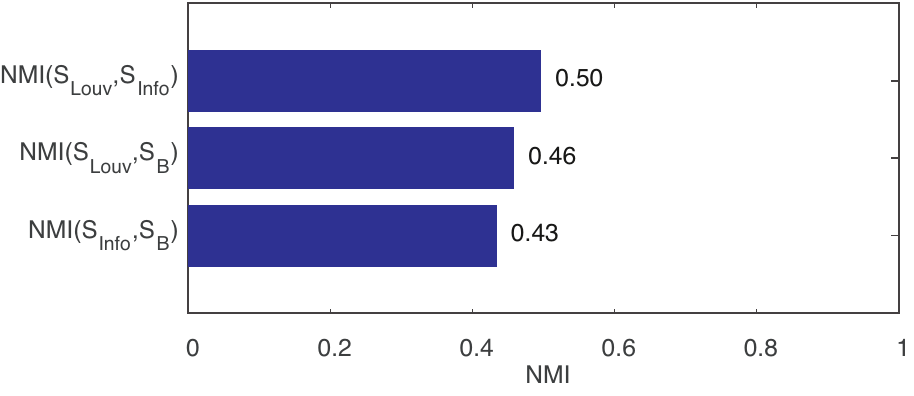}
	\caption{Pairwise NMI between $\mathbf{S}_{\text{Louv}}$, $\mathbf{S}_{\text{Info}}$, and $\mathbf{S}_{\text{B}}$ in the APS collaboration network.
	}
	\label{fig7}
\end{figure}

We now compare $\mathbf{S}_{\text{Louv}}$ and $\mathbf{S}_{\text{Info}}$ with a benchmark community assignment $\mathbf{S}_{\text{B}}$. We use PACS sections as the benchmark communities. To illustrate, subsections ``21. Nuclear Structure" and ``23. Radioactive decay and in-beam spectroscopy'' both belong to the benchmark layer community ``20. Nuclear Physics", whereas subsection ``1. Communication, education, history, and philosophy" belongs to the benchmark layer community ``0. General". We calculate that $\text{NMI}(\mathbf{S}_{\text{Louv}},\mathbf{S}_{\text{B}})$ and $\text{NMI}(\mathbf{S}_{\text{Info}},\mathbf{S}_{\text{B}})$ are about $0.45$ (see Fig.~\ref{fig7}), which suggests that the layer communities have a strong similarity (though are far from identical) to classification based on PACS sections.

\begin{figure*}[h]
\includegraphics[width= \textwidth]{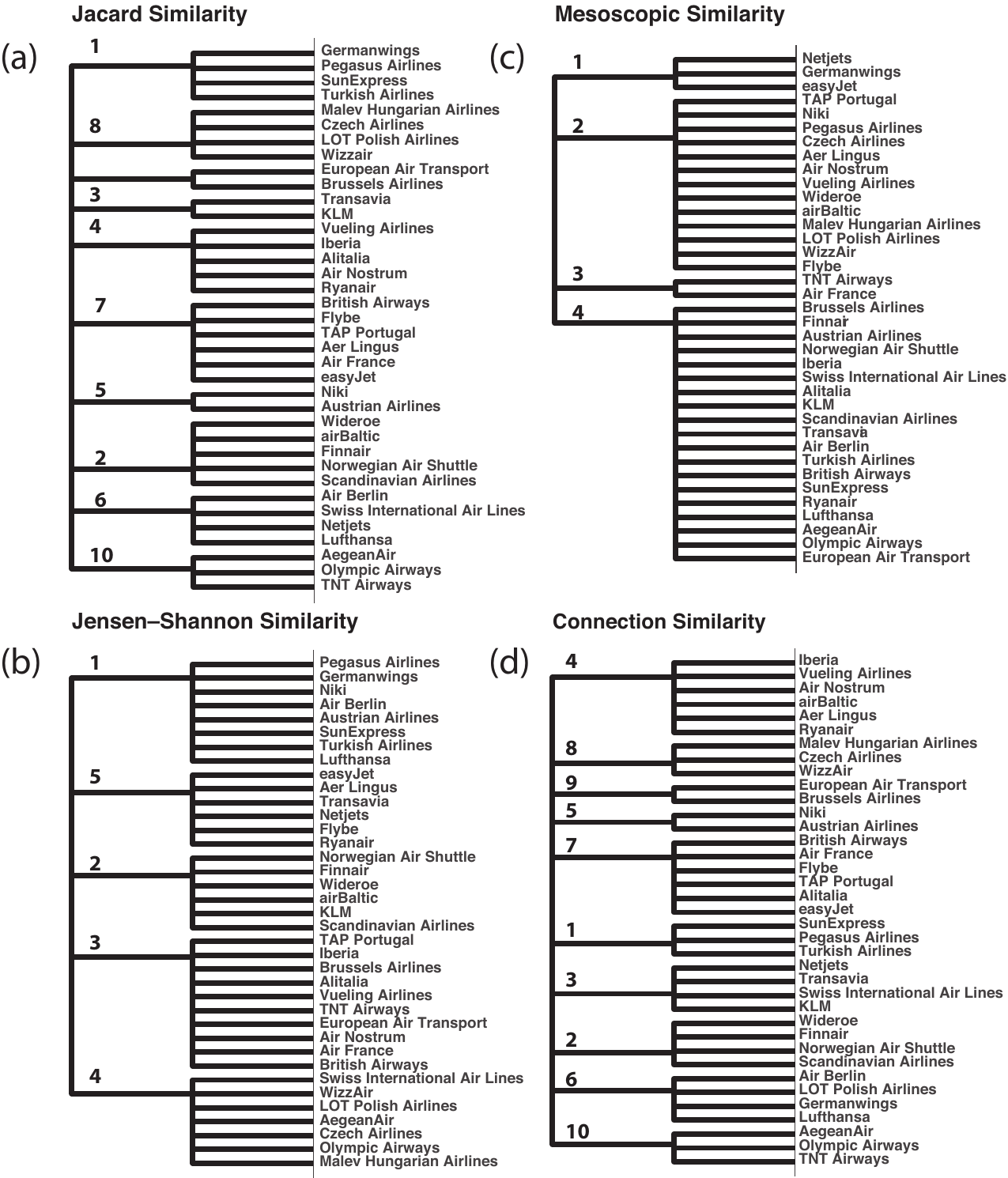}
	\caption{Airline layer communities that we find using (a) Jensen--Shannon (JS) distance, (b) the mesoscopic similarity indicator function, (c) Jaccard similarity, and (d) connection similarity. We detect these layer communities using the Louvain method \cite{blondel2008fast}. 
	}
	\label{fig8}
\end{figure*}

\begin{figure}
\includegraphics[width= \columnwidth]{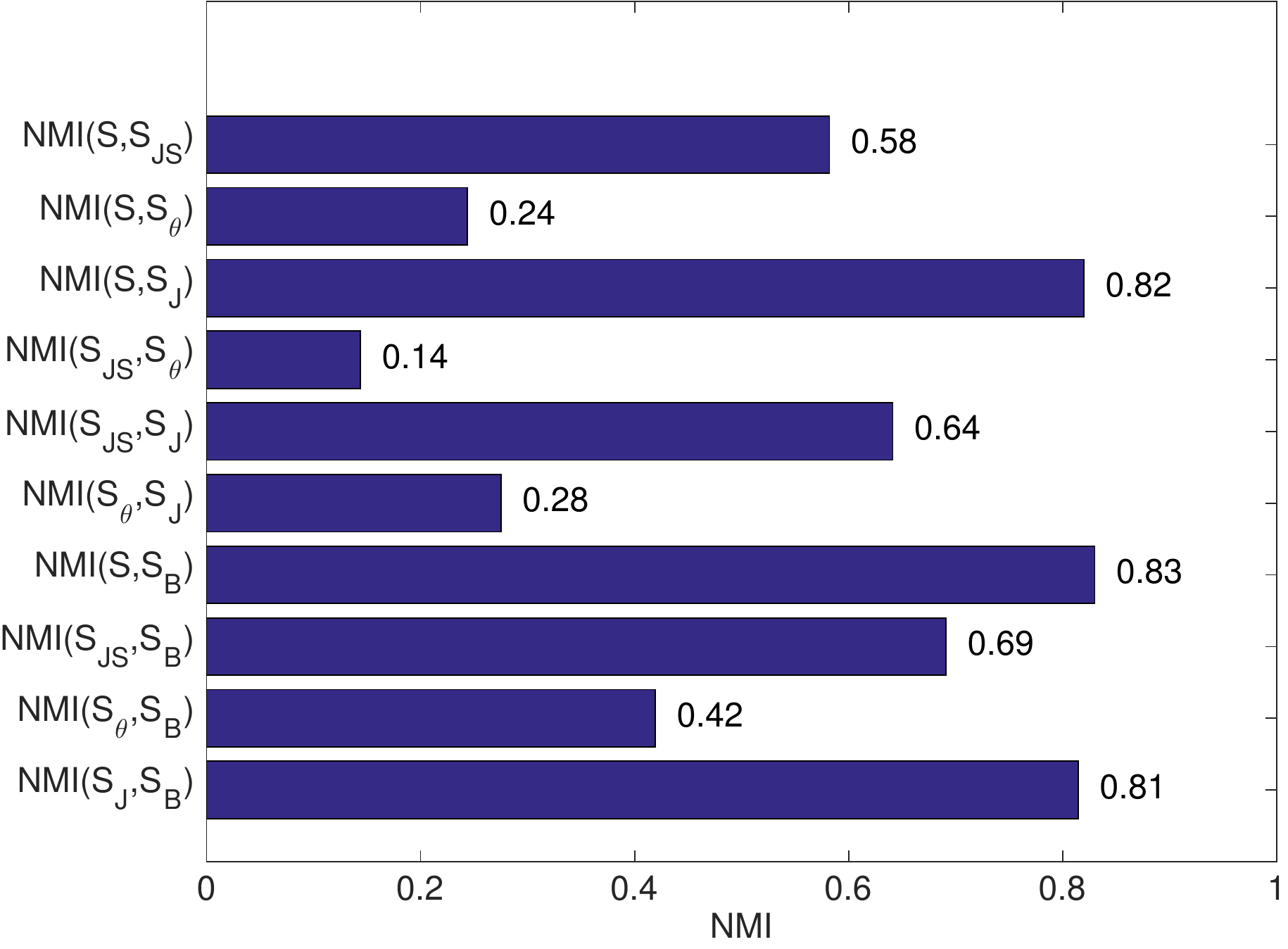}
	\caption{Pairwise NMI between $\mathbf{S}$, $\mathbf{S}_{\text{JS}}$, $\mathbf{S}_{\theta}$, $\mathbf{S}_{\text{B}}$, and $\mathbf{S}_{\text{J}}$ in the airline network.
	}
	\label{fig9}
\end{figure}


\subsubsection{Comparison with Jaccard, JS, and Mesoscopic Similarities} \label{3d}

We now compare connection similarity with Jaccard, JS \cite{de2015structural}, and mesoscopic \cite{iacovacci2016extracting} similarites by applying them to cluster airlines in the airline data set that we discussed in Section \ref{3b}. 

To calculate the mesoscopic similarity between layers, we first use the Louvain method to find node communities on each layer. We then use the algorithm of Iacovacci et al. \cite{MEMSA} to calculate mesoscopic similarity with respect to these node communities.

We then use the Louvain method \cite{blondel2008fast} to cluster airlines in the interlayer similarity matrices that we obtain from these similarity measures. We denote the layer communities that we find using connection similarity measure as $\mathbf{S}$, those that we find with the Jaccard similarity measure as $\mathbf{S}_{\text{J}}$, those that we find with the JS similarity measure as $\mathbf{S}_{\text{JS}}$, and those that we find with the mesoscopic similarity measure as $\mathbf{S}_{\theta}$. Following Section \ref{3b}, we construct a benchmark layer community $\mathbf{S}_{\text{B}}$ such that $S^{\alpha}_{\text{B}} = S^{\beta}_{\text{B}}$ if and only if airlines $\alpha$ and $\beta$ are based in the same country or in the same geographic region. 

In Fig.~\ref{fig8}, we show the airline communities that we obtain using the four different measures. In Fig.~\ref{fig9}, we plot the pairwise NMI between $\mathbf{S}$, $\mathbf{S}_{\text{J}}$, $\mathbf{S}_{\text{JS}}$, $\mathbf{S}_{\theta}$, and $\mathbf{S}_{\text{B}}$. The airline communities that we find using connection similarity are similar to those that we obtain using Jaccard similarity and JS similarity, but they are rather different from those that we find using mesoscopic similarity. We obtain an NMI between $\mathbf{S}$ and $\mathbf{S}_{\text{J}}$ of about $0.82$, an NMI between $\mathbf{S}$ and $\mathbf{S}_{\text{JS}}$ of about $0.58$, and an NMI between $\mathbf{S}$ and $\mathbf{S}_{\theta}$ of only about $0.24$.

We compare the airline communities that we find using connection, JS, and mesoscopic similarities with the benchmark community $\mathbf{S}_{\text{B}}$. We calculate that NMI($\mathbf{S}$,$\mathbf{S}_{B}$)$\approx0.83$ is larger than the other NMI values: NMI($\mathbf{S}_{\text{J}}$,$\mathbf{S}_{B}$)$ \approx 0.81$, NMI($\mathbf{S}_{\text{JS}}$,$\mathbf{S}_{B}$)$ \approx 0.69$, and NMI($\mathbf{S}_{\theta}$,$\mathbf{S}_{B}$)$\approx0.42$. Among these approaches, the airline communities that we find using connection similarity is most similar to the benchmark layer communities in the airline data set. 

We also calculate that NMI($\mathbf{S}$,$\mathbf{S}_{\text{JS}}$)$ \approx 0.58$, which suggests that the two approaches yield similar airline communities. This result is consistent with the findings of De Domenico et al. \cite{de2015structural}. When De Domenico et al. used a measure of JS distance to cluster layers with the aim of reducing the number of layers (and thus system size), they tended to combine layers with a large number of edge overlaps \cite{de2015structural}. Connection similarity quantifies layer similarity based directly on edge overlaps, so it is sensible that we find similar clusterings for connection and JS similarity.


\section{Conclusions} \label{four}

We proposed a new measure --- ``connection similarity'' --- to quantify similarity in connection patterns between two layers in a multiplex network. We used connect similarity to cluster layers in both synthetic and empirical multiplex networks. In the latter, we obtained layer communities that have real-world interpretations and are consistent with past studies. For example, our approach grouped airlines that are based in the same regions into layer communities in an airline  network.

Naturally, layer communities can differ when using different node-community detection algorithms (see Section \ref{3a}). For example, we found that InfoMap \cite{rosvall2008maps,MapCode} was unable to find the planted layer communities in our synthetic multiplex network, and it would be interesting in future work to explore benchmark multiplex networks with intricate interlayer dependencies \cite{bazzi2016generative,nicosia2013growing}

We proposed a measure of interlayer similarity based on edge overlaps, but there are also many other ways of measuring interlayer structural similarities \cite{nicosia2015measuring,battiston2014structural}, so in turn there are many ways of grouping layers into layer communities. Grouping approaches would benefit from further research into inherently multiplex structural measures to complement quantities like edge overlaps and interlayer degree correlations \cite{min2014network,lee2012correlated}. Moreover, because connection similarity does not take edge weights into account as a measure of layer similarity, it is also important to pursue layer similarity measures that take edge weights into account (e.g., a pairwise correlation coefficient of the edge weights in two layers \cite{mollgaard2016measure}). 


\begin{acknowledgements}

We thank participants in the University of Oxford's Networks Journal Club for helpful comments. 


\end{acknowledgements}






\end{document}